\documentclass{WileyMSP-template}

\usepackage{amsmath}

\begin{document}

\pagestyle{fancy}
%\rhead{\includegraphics[width=2.5cm]{vch-logo.png}}

\title{Robust magnetic order upon ultrafast excitation of an antiferromagnet}

\maketitle

% Author: Please give full first and last names for authors and include * after the name of all corresponding authors

\author{Sang-Eun Lee*}
\author{Yoav William Windsor}
\author{Alexander Fedorov}
\author{Kristin Kliemt}
\author{Cornelius Krellner}
\author{Christian Schü\ss ler-Langeheine}
\author{Niko Pontius}
\author{Martin Wolf}
\author{Unai Atxitia}
\author{Denis V. Vyalikh}
\author{Laurenz Rettig*}

% Dedication

\dedication{}

% Affiliations: Please provide adacemic titles (Prof. or Dr.) for all authors where applicable, and include an institutional email address for all corresponding authors
\begin{affiliations}
Sang-Eun Lee\\
Department of Physical Chemistry, Fritz-Haber-Institut der Max-Planck-Gesellschaft, Faradayweg 4-6, 14195 Berlin, Germany\\
Email Address: lee@fhi-berlin.mpg.de\\

Dr. Yoav William Windsor\\
Department of Physical Chemistry, Fritz-Haber-Institut der Max-Planck-Gesellschaft, Faradayweg 4-6, 14195 Berlin, Germany\\

Dr. Alexander Fedorov\\
Institute for Solid State Research, Leibniz IFW Dresden, Helmholtzstr. 20, 01069 Dresden, Germany\\

Dr. Kristin Kliemt\\
Physikalisches Institut, Goethe-Universität Frankfurt, Max-von-Laue-Str. 1, 60438 Frankfurt am Main, Germany\\

Prof. Dr. Cornelius Krellner\\
Physikalisches Institut, Goethe-Universität Frankfurt, Max-von-Laue-Str. 1, 60438 Frankfurt am Main, Germany\\

Dr. Christian Schü\ss ler-Langeheine\\
Helmholtz-Zentrum Berlin für Materialien und Energie GmbH, Albert-Einstein-Str. 15, 12489 Berlin, Germany\\

Dr. Niko Pontius\\
Helmholtz-Zentrum Berlin für Materialien und Energie GmbH, Albert-Einstein-Str. 15, 12489 Berlin, Germany\\

Prof. Dr. Martin Wolf\\
Department of Physical Chemistry, Fritz-Haber-Institut der Max-Planck-Gesellschaft, Faradayweg 4-6, 14195 Berlin, Germany\\

Dr. Unai Atxitia\\
Dahlem Center for Complex Quantum Systems and Fachbereich Physik, Freie Universität Berlin, Arnimallee 14, 14195 Berlin, Germany\\

Prof. Dr. Denis V. Vyalikh\\
Donostia International Physics Center (DIPC), Paseo Manuel de Lardizabal 4, 20018 Donostia/San Sebastián, Basque Country, Spain\\
IKERBASQUE, Basque Foundation for Science, 48009 Bilbao, Spain\\

Dr. Laurenz Rettig\\
Department of Physical Chemistry, Fritz-Haber-Institut der Max-Planck-Gesellschaft, Faradayweg 4-6, 14195 Berlin, Germany\\
Email Address: rettig@fhi-berlin.mpg.de\\
\end{affiliations}

% Keywords: Please provide a minimum of three and a maximum of seven keywords, separated by commas


% Abstract should be written in the present tense and impersonal style (i.e., avoid we), and be at most 200 words long
\begin{abstract}
The ultrafast manipulation of magnetic order due to optical excitation is governed by the intricate flow of energy and momentum between the electron, lattice and spin subsystems. While various models are commonly employed to describe these dynamics, a prominent example being the microscopic three temperature model (M3TM), systematic, quantitative comparisons to both the dynamics of energy flow and magnetic order are scarce. Here, we apply a M3TM to the ultrafast magnetic order dynamics of the layered antiferromagnet GdRh$_{2}$Si$_{2}$. The femtosecond dynamics of electronic temperature, surface ferromagnetic order, and bulk antiferromagnetic order were explored at various pump fluences employing time- and angle-resolved photoemission spectroscopy and time-resolved resonant magnetic soft x-ray diffraction, respectively. After optical excitation, both the surface ferromagnetic order and the bulk antiferromagnetic order dynamics exhibit two-step demagnetization behaviors with two similar timescales ($<$1 ps, $\sim$10 ps), indicating a strong exchange coupling between localized 4f and itinerant conduction electrons. Despite a good qualitative agreement, the M3TM predicts larger demagnetization than our experimental observation, which can be phenomenologically described by a transient, fluence-dependent increased Néel temperature. Our results indicate that effects beyond a mean-field description have to be considered for a quantitative description of ultrafast magnetic order dynamics.
\end{abstract}

% Text: Please use section headings and subheadings as specified below. For communications, all section headings apart from Experimental Section should be removed
% Please make the first reference to a display item bold: \textbf{Figure 1}
% Do not abbreviate Figure, Equation, etc.; display items are always singular, i.e., Figure 1 and 2.
% Equations are always singular, i.e., Equation 1 and 2, and should be inserted using the {equation} environment, not as graphics
% Please do not use footnotes in the text, additional information can be added to the Reference list.

\section{Introduction}\label{Introduction}
\begin{figure*}[ht]
\includegraphics[width=\linewidth]{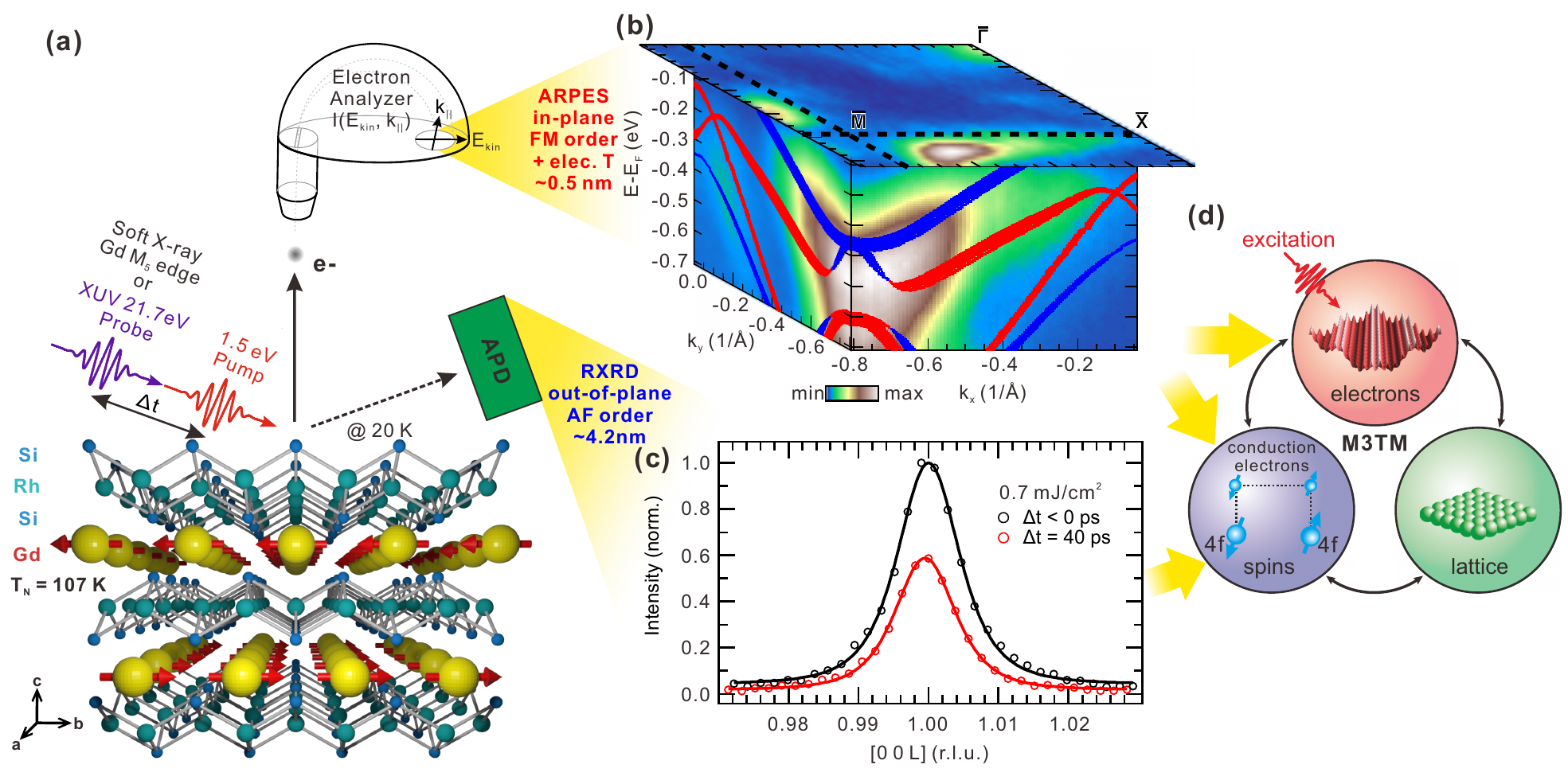}
\caption{(a) Crystal structure of GdRh$_2$Si$_2$ and sketch of experimental setups. GdRh$_2$Si$_2$ is a quasi-2D intermetallic material with a layered tetragonal crystal structure of ThCr$_2$Si$_2$ type (a = b = 4.03 Å, c = 9.88 Å). Below 107 K, ferromagnetic Gd layers separated by the Si-Rh-Si blocks couple antiferromagnetically. The experiments were conducted at 20 K unless specified. (b) Volumetric representation of the ARPES intensity around the $\overline{\textrm{M}}$ point of the surface Brillouin zone of the Si-terminated surface of GdRh$_2$Si$_2$. The constant energy contour at E-E$_F$ = 0 shows the Fermi surface topology (integration width = 7.7 meV). The direction of the k$_x$ and k$_y$ cuts along the $\overline{\textrm{M}}$-$\overline{\textrm{X}}$ directions are indicated by dashed lines. A density functional theory (DFT) calculation of the spin-resolved surface state band structure is overlaid \cite{guttler_robust_2016}. Majority (minority) spin states are shown as red (blue) ribbons. (c) (001) magnetic diffraction peak of GdRh$_2$Si$_2$ measured with trRXD. 40 ps after excitation (red), the diffraction intensity is suppressed compared to the diffraction intensity before t$_0$ (black). (d) Schematics of the M3TM for lanthanide-based antiferromagnets. The M3TM takes the electronic temperature dynamics from trARPES measurement as an input to predict magnetic order dynamics of the itinerant conduction electrons and the localized 4f magnetic moments, which will be compared with the experimental results from (b) trARPES and (c) trRXD measurements in Section \ref{Analysis}.}\label{F1}
\end{figure*}

Since the discovery of ultrafast demagnetization by Beaurepaire et al. \cite{beaurepaire_ultrafast_1996}, numerous studies have applied variations of three-temperature models (3TM) to describe experimental ultrafast magnetization dynamics \cite{koopmans_explaining_2010,shim_role_2020,atxitia_controlling_2014,chen_ultrafast_2019,frietsch_role_nodate,sultan_electron-_2012,radu_transient_2011,frietsch_disparate_2015,atxitia_evidence_2010}. By introducing effective temperatures for the transient electronic, lattice and spin degrees of freedom (see Fig. \ref{F1}-d), the 3TM provides an intuitive, phenomenological approach for the quantitative analysis of ultrafast magnetization dynamics using three coupled differential equations to describe the mutual energy transfer between the subsystems. The microscopic three-temperature model (M3TM) improved the 3TM by considering momentum conservation during the ultrafast magnetization dynamics via the Elliott-Yafet spin-flip scattering substituting a phenomenological spin temperature with magnetization \cite{koopmans_explaining_2010}. Such formulations are related to the Landau-Lifshitz-Bloch (LLB) equation, where the specifics of the couplings to the electrons and phonons in magnets are encoded in the damping parameter of the macrospin dynamics \cite{atxitia_evidence_2010,atxitia_fundamentals_2016,atxitia_ultrafast_2011,nieves_quantum_2014}. M3TMs also explain the two-step demagnetization behavior typically found in lanthanides, such as Gd, due to enhanced spin-flip scattering within a non-thermal system \cite{koopmans_explaining_2010}. However, as these materials are characterized by indirect RKKY exchange interaction between localized 4f electrons with large magnetic moments, mediated through weakly spin-polarized conduction electrons, distinct dynamics of these magnetic subsystems also needs to be considered, and has been recently reported for Gd and Tb \cite{frietsch_role_nodate}.

As both, the M3TM and the LLB equation calculate the magnetization dynamics using the ultrafast transient temperature changes of the subsystems as an input, in order to test the accuracy of such models, a rigorous comparison to experimental values for the temperature and magnetization dynamics measured under comparable conditions is required. 
However, a quantitative determination of such ultrafast temperature transients is challenging to achieve experimentally \cite{zahn_lattice_2021,tengdin_critical_nodate,andres_separating_2015,andres_role_2021,gort_ultrafast_2020,perfetti_ultrafast_2007,vasileiadis_ultrafast_2018,waldecker_electron-phonon_2016}. Furthermore, a simultaneous measurement of the temperature and magnetization dynamics is typically not feasible, and combining different experiments imposes the difficulty of ensuring identical experimental conditions. Thus, only few quantitative comparisons of measured transient temperature and ultrafast magnetization dynamics are available \cite{beaurepaire_ultrafast_1996,zahn_lattice_2021,tengdin_critical_nodate,andres_separating_2015,andres_role_2021,gort_ultrafast_2020}. The electron and lattice temperature are the most relevant dynamical input parameters describing the energy relaxation in the M3TM, however, studies using experimental electronic temperature or phonon dynamics as input for modelling the experimental magnetization dynamics are scarce, in particular systematic studies for various excitation fluences. A recent study of Zahn et al. \cite{zahn_lattice_2021} used a model to calculate the atomistic spin dynamics based on the stochastic Landau-Lifshitz-Gilbert equation, the atomistic counterpart of the LLB equation, and compared the results to electron diffraction data and the literature values of magneto-optical Kerr effect measurements from Ref. \cite{you_revealing_2018}. Remarkably, even for the single fluence analyzed in that study, the model simulations based on the measured ultrafast lattice dynamics showed significant deviations to the experimentally measured magnetization dynamics.

While the majority of studies so far has concentrated on ferromagnets, antiferromagnetic (AF) materials recently have attracted strong attention due to their potential use in low-power-consumption and high-stability next-generation memory devices \cite{jungwirth_antiferromagnetic_2016, thielemann-kuhn_ultrafast_2017, bai_functional_2020}. In addition, the lack of a net magnetic moment in AF materials promises faster manipulation of magnetic order. However, such an arrangement of compensating spin-sublattices also poses additional challenges, as new experimental approaches to control and interact with magnetic order are required. Therefore, while dynamics in ferromagnetic materials has been studied extensively, experimental studies of ultrafast spin dynamics in AF materials are still scarce. On the theory side, the compensating alignment of their magnetic sublattices allows for additional scattering channels, such as inter-sublattice exchange of angular momentum \cite{windsor_exchange_2022,thielemann-kuhn_ultrafast_2017} which are considered in more recent M3TMs based on the LLB equation \cite{atxitia_fundamentals_2016,jakobs_universal_2022,jakobs_exchange-enhancement_2022} applied in our study. 

Here, we investigate the ultrafast magnetic order dynamics and transient electronic temperature in the layered quasi-2D antiferromagnet, GdRh$_2$Si$_2$ using time- and angle-resolved photoelectron spectroscopy (trARPES) and time-resolved resonant magnetic soft x-ray diffraction (trRXD) (see Experimental Section and Fig. \ref{F1}-a). GdRh$_2$Si$_2$ is a 4f antiferromagnet, where Gd atomic layers are separated by strongly bonded Si-Rh-Si blocks along the [001] direction as shown in Fig. \ref{F1}-a \cite{kliemt_single_2015}. Due to the localized nature of the 4f moments, magnetic order is mediated through the spin-polarized itinerant Gd $d$, Si and Rh conduction electrons via the indirect RKKY exchange interaction. Our choice of experimental methods allows us to get a full picture on the ultrafast dynamics of both localized and itinerant magnetic order, as well as the electronic temperature evolution after excitation: Surface-sensitive trARPES allows for the simultaneous analysis of the magnetization-dependent transient exchange splitting of a Si-derived surface state and of the time-dependent electron distribution function (Fig. \ref{F1}-b). It thereby provides the unique opportunity to study the ultrafast dynamics of both the electronic temperature and the in-plane surface magnetization of itinerant conduction electrons within a single experiment. Additionally, bulk-sensitive trRXD is used to study directly the temporal evolution of long-range, out-of-plane AF order of the localized Gd 4f moments (Fig. \ref{F1}-c). In Section \ref{Experimental Results}, we present the above-mentioned experimental results. In Section \ref{Analysis}, by applying a suitable M3TM for AF systems based on the LLB equation, we \emph{quantitatively} compare the model’s prediction using the measured electronic temperature as input to the magnetic order dynamics of both the itinerant conduction electrons and the localized 4f moments from trARPES and trRXD, respectively, as schematically shown in Fig. \ref{F1}-d. While the model allows for a good qualitative description of the ultrafast magnetic order dynamics, we show in Section \ref{Discussion} that for increasing excitation fluence, the material shows an increased robustness against demagnetization. Phenomenologically, we can describe this behavior with a transiently enhanced Néel temperature. Additionally, the initial, sub-ps demagnetization significantly exceeds the model prediction, suggesting an enhanced inter-sublattice momentum transfer rate in the non-thermal system.

\section{Experimental Results}\label{Experimental Results}
\begin{figure*}[ht]
\includegraphics[width=\linewidth]{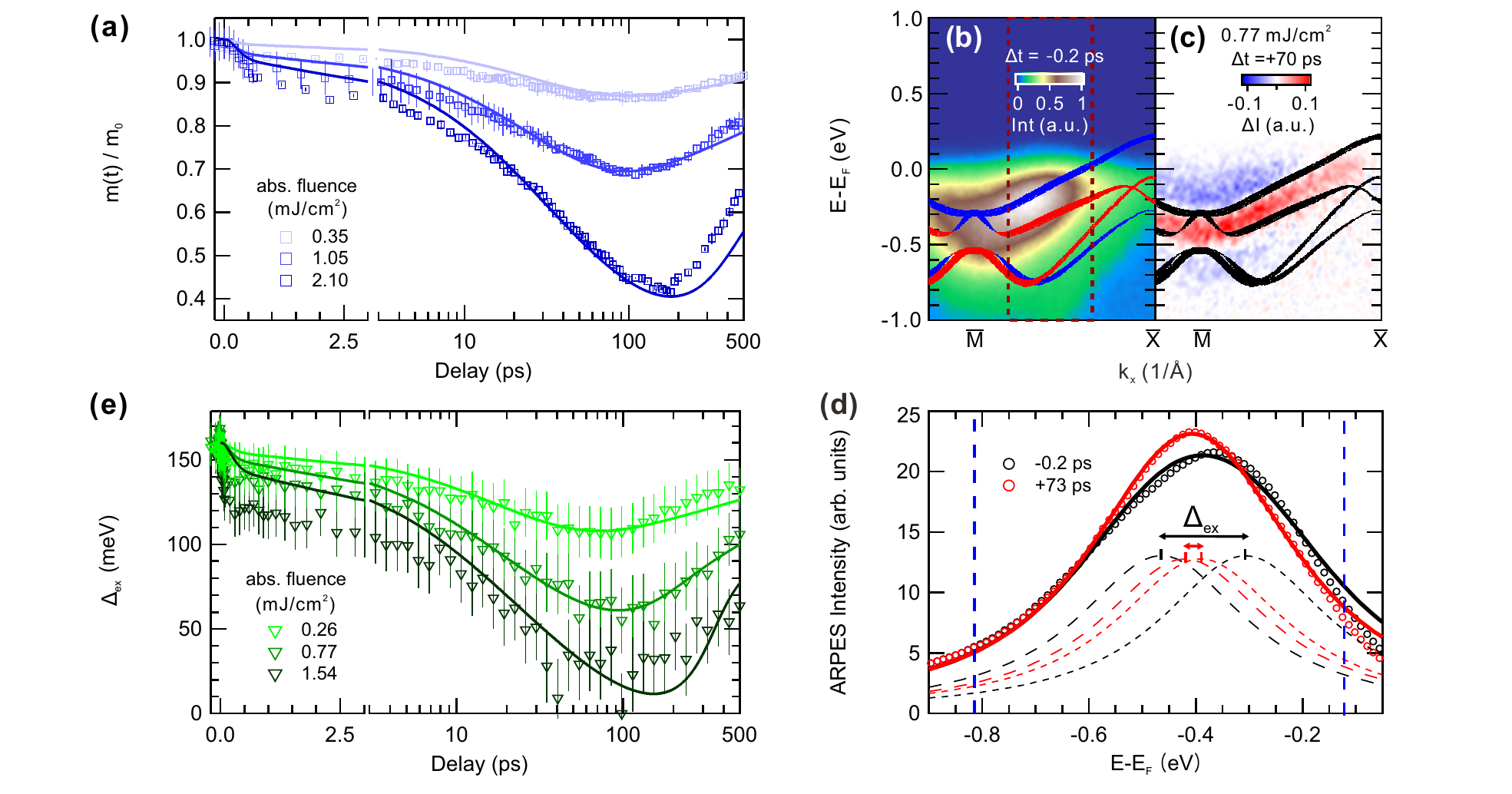}
\caption{(a) Ultrafast dynamics of the normalized (001) magnetic diffraction amplitude (squares). Solid lines show the simulated magnetization dynamics by the M3TM (see Section \ref{Analysis}) taking the probe depth of RXD into account. Note that the second half of the time axis is log-scaled. (b) ARPES intensity at -0.2 ps and (c) its intensity difference upon pump excitation at 70 ps delay normalized by the maximum intensity at -0.2 ps. The ARPES intensity is homogeneously increased along the dispersion of the surface state after excitation. DFT calculations of the spin-resolved (red: Majority state, blue: Minority state) and spin-integrated (black) surface state from Ref. \cite{guttler_robust_2016} are overlaid. (d) EDCs of the momentum-integrated surface state (energy-corrected for its dispersion, see text) at -0.2 ps and 70 ps (open circles) modeled by two Lorentzian profiles convolved with a Gaussian instrument response function (thick solid lines). The transient exchange splitting ($\Delta_{ex}$), extracted from the distance between the two peaks (thin dashed lines: Majority spin state, thin dotted lines: Minority spin state), decreases after excitation. The blue dotted lines mark the region of interest used for extracting the exchange splitting. (e) Ultrafast dynamics of the exchange splitting (triangles). Error bars are confidence interval of exchange splitting extraction detailed in Supplementary Information Section A. Solid lines describe the simulated magnetization dynamics by the M3TM (see Section \ref{Analysis}) taking the probe depth of trARPES into account, scaled to the exchange splitting at 19 K (160 meV) \cite{guttler_robust_2016}. Note that the second half of the time axis is log-scaled. Note that the figures are placed in clockwise order.}\label{F2}
\end{figure*}
\subsection{Femtosecond dynamics of long-range Gd 4f antiferromagnetic ordering}
Using resonant magnetic x-ray diffraction we measured the response to photoexcitation of the (001) intensity of GdRh$_2$Si$_2$. The femtosecond dynamics of the (001) diffraction peak amplitude recorded at constant momentum transfer Q is shown in Fig. \ref{F2}-a for selected pump fluences (squares). Here, the peak amplitudes have been separated from a transient reorientation of the magnetic structure based on a procedure combining several azimuthal orientations, as detailed in Ref. \cite{windsor_deterministic_2020}. The normalized diffraction amplitude dynamics exhibits two-step decay according to a biexponential fit with time constants of a $<$ 1 ps, and a $\sim$10 ps \cite{windsor_deterministic_2020}, as commonly observed in lanthanide magnets, followed by a slow recovery after $\sim$100 ps.

\subsection{Exchange splitting dynamics}
Next, we used trARPES to study the photo-induced evolution of a Si-derived surface state residing at the large projected band gap at the $\overline{\textrm{M}}$ point (Fig. \ref{F1}-b). In GdRh$_2$Si$_2$, the localized Gd 4f electrons predominantly carry the magnetic moments, and the conduction electrons from Rh, Si and Gd 5d6s mediate the RKKY interaction between the Gd layers \cite{kliemt_single_2015}\cite{windsor_exchange_2022}. In the AF state, the surface state exhibits a sizeable exchange splitting, which is mediated via RKKY exchange coupling to the localized Gd 4f moments from the sub-surface in-plane FM Gd layer \cite{guttler_robust_2016}. The exchange splitting sets in at 90 K, notably lower than the bulk Néel temperature $T_N$ = 107 K, and reaches $\sim$160 meV at 19 K \cite{guttler_robust_2016}.

The trARPES intensity along the $\overline{\textrm{M}}$-$\overline{\textrm{X}}$ direction is shown in Fig. \ref{F2}-b. Due to the limited energy resolution of our trARPES setup (150 meV), which is of similar magnitude as the exchange splitting, the exchange-split bands of the surface state are difficult to resolve and appear as a single dispersing band. The transient trARPES intensity along this cut was measured for various pump-probe delays. The pump-induced difference $\Delta I/I$ = $[I\left(70 ps\right)$ $-$ $I\left(-0.2 ps\right)]/I_{max.}\left(-0.2 ps\right)$ is shown in Fig. \ref{F2}-c, and exhibits a narrowing of the surface state profile, consistent with a decrease of the exchange splitting.

In order to extract the transient exchange splitting, we analyze the transient energy distribution curves (EDCs) of the surface state using an empirical model fit as described in the following. As the exchange splitting is almost constant in a large range along the linear dispersion of the surface state, we integrate the EDCs along the $\overline{\textrm{M}}-\overline{\textrm{X}}$ direction by correcting for the peak’s energy-momentum dispersion (see Supplementary Information Section A for details of this procedure). The exchange splitting has been extracted by fitting the dispersion-corrected, integrated EDC with two Lorentzian profiles representing the two spin-split surface states, convolved with a Gaussian accounting for the energy resolution (Figure 2-d). Its width is determined from the EDC at $T=150$ K (Fig. A.1 in the Supplementary Information) where the exchange splitting vanishes \cite{guttler_robust_2016}. The exchange splitting dynamics at various fluences extracted from this procedure are shown in Figure 2-e (triangles), and the exchange splitting before excitation is found in agreement with published results \cite{guttler_robust_2016}. Similar to the dynamics of the normalized diffraction amplitude, the exchange splitting dynamics exhibits a two-step  demagnetization ($<$1 ps and $\sim$10 ps \cite{windsor_deterministic_2020}), followed by a slow recovery after $\sim$100 ps. 

\subsection{Electronic temperature dynamics}
\begin{figure}[ht]
\centering
\includegraphics{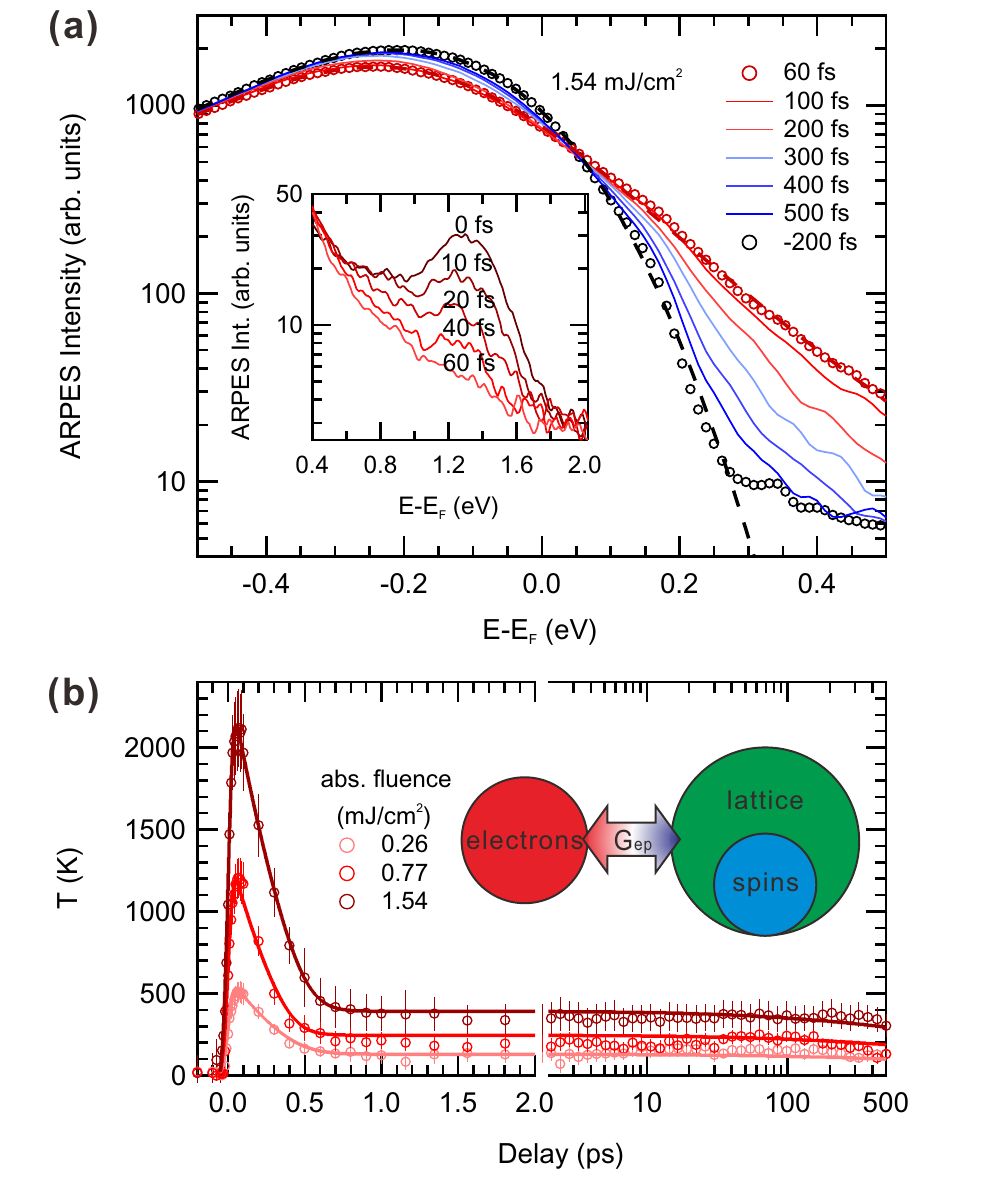}
\caption{(a) EDCs integrated at the Fermi momentum (red dashed box in Fig. \ref{F2}-b) for various pump-probe delays on a logarithmic intensity scale. Dashed lines are fits to a model function (see text) plotted for the EDCs at -200 fs and 60 fs. (Inset) Evolution of the transient trARPES intensity of an unoccupied state during the thermalization time of the system. (b) Ultrafast electronic temperature dynamics (circles). Solid lines show the electronic temperature dynamics by the 2TM (see Section \ref{Analysis}) taking the probe depth of trARPES into account. The 2TM is schematically described by a diagram in the corner. Note that the second half of the time axis is log-scaled.}\label{F3}
\end{figure}
The electronic temperature is extracted from the transient trARPES intensity evolution integrated around the Fermi momentum $k_F$ (red dashed lines in Fig. \ref{F2}-b). Fig. \ref{F3}-a shows EDCs integrated at $k_F$ for various pump probe delays on a logarithmic intensity scale. The sharp drop-off at E$_F$ is due to the Fermi-Dirac distribution function, which encodes the transient electronic temperature. In order to quantify the change of the electronic temperature as function of delay, we have modeled the EDC by a phenomenological density-of-states function consisting of a Lorentzian profile and constant background multiplied with the Fermi-Dirac distribution function, and convolved with the instrumental response function. Fig. \ref{F3}-a shows exemplary fits to the data at -200 fs and +60 fs, which describes the data very well especially the Fermi-edge region (E-E$_F$ $\leq$ $\pm$0.3 eV), which is relevant for extracting the electronic temperature. At early times ($<$60 fs), there are some deviations at energies above E-E$_F$ $>$ +0.3 eV, originating from non-thermalized electrons right after excitation, which do not follow a Fermi-Dirac distribution. In particular, we also observe the transient occupation of an electronic state at +1.2 eV above E$_F$, which decays on a timescale of $\sim$60 fs (see inset of Fig. \ref{F3}-a). Subsequently, the system is thermalized and is well described by fits to the Fermi-Dirac function. At pump fluences larger than 1 mJ/cm$^2$, the EDCs are influenced by space charge effects \cite{oloff_pump_2016} leading to a time-dependent shift of the Fermi energy. Therefore, a time-dependent correction of the Fermi energy has been applied as detailed in Supplementary Information Section B. The extracted transient electronic temperatures at various pump fluences are shown in Fig. \ref{F3}-b. A detailed discussion and comparison of the fits and fitting results can be found in Supplementary Information Section C.

At all pump fluences, the extracted effective electronic temperature steeply increases within the first 50 fs due to the absorption of the pump pulse energy, subsequently decreases within $\sim$0.5 ps via redistribution of energy to the lattice governed by electron-phonon (e-ph) coupling, and finally slowly recovers to the starting temperature within several 100 ps by heat diffusion (Fig. \ref{F3}-b).

\section{Analysis}\label{Analysis}
Our experimental data consist of the ultrafast dynamics of the electronic temperature, the exchange splitting, and the normalized (001) magnetic diffraction amplitude upon 1.55 eV pump excitation. The exchange splitting and the diffraction amplitude dynamics exhibit a very similar two-step demagnetization ($<$1 ps, $\sim$10 ps) and subsequent recovery after $\sim$100 ps, suggestive of a common physical origin. The electronic temperature also exhibit dynamics on similar timescales. In order to consistently describe our experimental results, we modeled the transient electronic temperature and demagnetization dynamics using an M3TM based on the LLB equation, modified to account for AF angular momentum exchange \cite{atxitia_fundamentals_2016,atxitia_ultrafast_2011,jakobs_universal_2022,jakobs_exchange-enhancement_2022,windsor_exchange_2022}.

Within this model, the electronic temperature $T_e$, the lattice temperature $T_p$, and the magnetization $m$ are described by three coupled differential equations:

\begin{eqnarray}
C_e\frac{dT_e}{dt}=&G_{ep}\left( T_p-T_e \right)+\nabla \left( k_e \nabla_z T_e\left(z \right)\right) + S\left(z,t\right)&\label{T_e}\\
C_p\frac{dT_p}{dt}=&-G_{ep}\left(T_p-T_e\right)&\label{T_p}\\
\frac{dm}{dt}=&Rm\frac{T_p}{T_N}\left(1-\frac{m}{B_{7/2}\left(m E_{ex}/k_B T_e\right)}\right)\left(1+\frac{2}{z_{n.n.}m}\right)&\label{m}
\end{eqnarray}

In Equation \eqref{T_e}, $C_e = \gamma_0 T_e$ is the electronic heat capacity, where $\gamma_0$ is the Sommerfeld coefficient. $G_{ep}$ is the electron-phonon coupling constant. $k_e$ is the electronic thermal conductivity, and $S(z,t)$ models the depth- and time-dependent pump excitation given by a Gaussian distribution of the pump pulse width and its exponential suppression according to the pump pulse penetration depth of 20.6 nm \cite{windsor_deterministic_2020}. In Equation \eqref{T_p}, $C_p$ is the lattice heat capacity. As we see a predominant reduction of magnetic order on a timescale comparable to lattice heating, and the localized nature of the Gd 4f moments, we include the spin heat capacity in the lattice heat capacity (Fig. \ref{F3}-b). The lattice heat capacity is taken from the specific heat of LuRh$_2$Si$_2$ (a paramagnetic (PM) sister compound of AF GdRh$_2$Si$_2$ due to the fully-filled 4f orbital of the Lu ions) and the spin heat capacity extracted as the difference of the specific heat between AF GdRh$_2$Si$_2$ and PM LuRh$_2$Si$_2$ as detailed in Supplementary Information Section D \cite{kliemt_single_2015}. In Equation \eqref{m}, $R=(8a_{sf} G_{ep} \mu_B k_B T_{N}^{2})/(\mu_{at} E_{D}^{2})$ is a material-specific factor proportional to $a_{sf}$, the spin-flip scattering probability, and $T_N$ is the Néel temperature. $B_{7/2}(E_{ex}/k_B T_e)$ is the Brillouin function, where $E_{ex}$ is the exchange energy (proportional to the Néel temperature of AF GdRh$_2$Si$_2$), and k$_B$ is the Boltzmann constant. $\mu_{at}$, $E_D$ are the atomic magnetic moment of Gd and the Debye energy, $k_B T_D$, respectively. The Debye temperature $T_D$ is estimated by fitting the Debye model to the lattice heat capacity. The term $2/(z_{n.n.} m)$ in Equation \eqref{m} describes the antiferromagnetic angular momentum transfer between different Gd 4f sublattices, where z$_{n.n.}$ = 8 is the number of AF coupled nearest neighbors. We note here that while the model considers the response of a bulk-coordinated system, the reduced magnetic coordination at the surface will cause slightly larger demagnetization  for a given pump excitation compared to a pure bulk system. As will be shown below, the model predicts much larger demagnetization even in the bulk limit. Therefore, we do not consider surface effects in the AF coordination in this study.

\begin{table}[ht]
\centering
\begin{tabular}{c|c||c|c}
$\gamma_0$ (J/K$^{2}$m$^{3}$) & 247-760 & $T_D$ (K) & 430  \\
\hline
k$_e$ (W/m K)    & 0.5  & G$_{ep}$ (J/K s m$^{3}$) & (8.5-13)$\times$10$^{17}$\\
\hline
$T_N$ (K)    &\begin{tabular}[c]{@{}c@{}}130-340 (ex. split)\\265-600 (001 amp.)\end{tabular}  & R (1/ps)  & \begin{tabular}[c]{@{}c@{}}0.074 (ex. split)\\0.055 (001 amp.)\end{tabular}   \\
\hline
E$_{ex}$/$T_N$ (eV/K)   & 0.002  & $\mu_{at}$ ($\mu_B$)  & 7.55  
\end{tabular}
\caption{Physical parameters of GdRh$_2$Si$_2$ for the 2TM and the M3TM.}\label{T1}
\end{table}

Equations \eqref{T_e} and \eqref{T_p} (the standard two-temperature model) describe the energy flow from the electrons, which are heated by the source term $S$, into the lattice and the heat transport due to diffusion \cite{lisowski_ultra-fast_2004}. Equation \eqref{m} is derived from the LLB equation, which is extended to antiferromagnets and combined with the M3TM \cite{jakobs_exchange-enhancement_2022}. It describes the magnetization dynamics depending on $T_e$ and $T_p$. In order to account for the different probe depths of the two probes (trARPES: $\sim$0.5 nm, trRXD: $\sim$4.2 nm), simulations are performed as function of depth $z$, and weighted with the respective probe depths. Simulated time evolutions of $T_e$, $T_p$ and $m$ at selected absorbed pump fluences are overlaid on the experimental results (Fig. \ref{F2}-a/e, \ref{F3}-b, and \ref{F4}-a/b, \ref{F5}-a). For determining the model parameters, at a given fluence, first a numerical solution of Equations \eqref{T_e} and \eqref{T_p} is fit to the electronic temperature dynamics (Fig. \ref{F3}-b), and subsequently a numerical solution of Equation \eqref{m} is fit to the magnetization dynamics of both, the (001) magnetic diffraction amplitude, and the surface state exchange splitting (Fig. \ref{F2}). While the model can describe the qualitative evolution of the curves very well, we found that for a quantitative description of the fluence dependent results, we need to vary a number of model parameters significantly. The physical parameter ranges used for the simulation are listed in Table \ref{T1}. 

\section{Discussion}\label{Discussion}
In this section we will discuss the M3TM simulations, and the fluence dependence of the extracted parameters. First, we will compare the dynamics at a given fluence, and then discuss the fluence dependence of the results.

\subsection{Comparison of electronic temperature and magnetic order dynamics}
\begin{figure}[ht]
\centering
\includegraphics{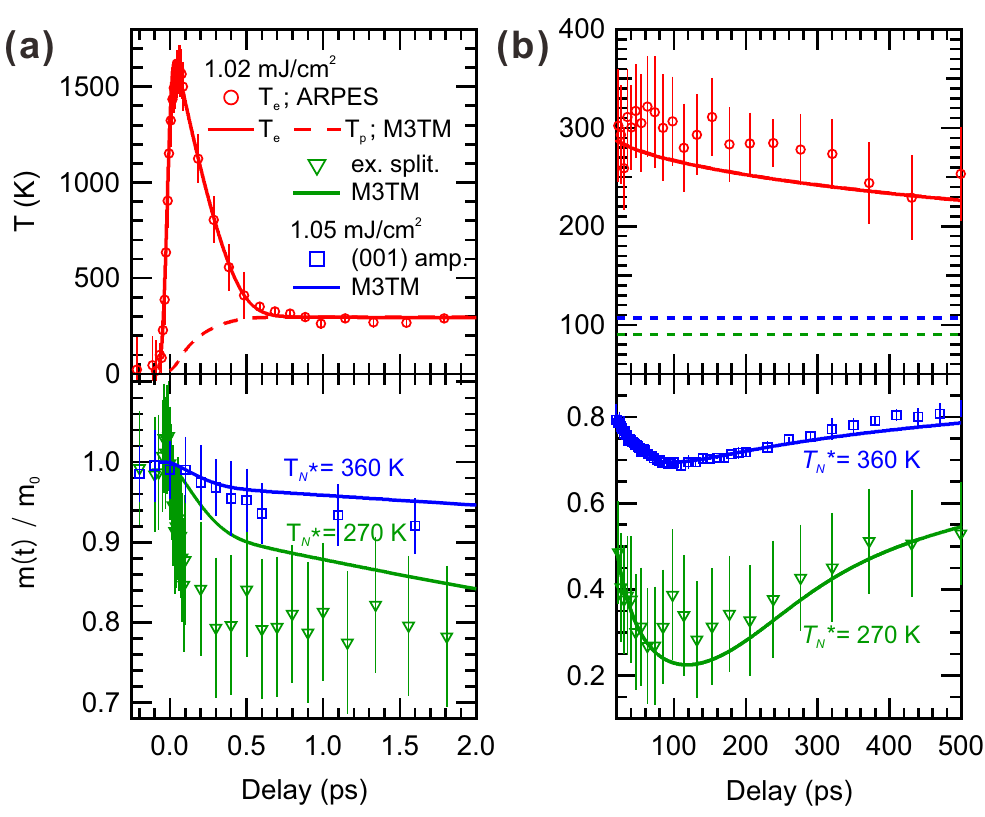}
\caption{(a) Short-term and (b) long-term dynamics of the electronic temperature (red circles), the normalized exchange splitting (green triangles) and the (001) diffraction amplitude (blue squares). Red solid (dashed) lines are electronic (lattice) temperature dynamics simulated by the M3TM. Blue and green solid lines show the simulated magnetization dynamics taking the probe depth of trRXD and trARPES into account, respectively. Dashed lines in (b) indicate the reference sample $T_N$ (blue: bulk AF order \cite{windsor_deterministic_2020}; green: exchange splitting \cite{guttler_robust_2016}).}\label{F4}
\end{figure}
Fig. \ref{F4}-a/b shows the ultrafast dynamics of the electronic temperature dynamics together with the dynamics of the normalized magnetic order parameter of both itinerant (surface state exchange splitting) and localized (trRXD amplitude) magnetic order within the first two ps (Fig. \ref{F4}-a), and on a longer timescale (Fig. \ref{F4}-b) for an absorbed fluence of 1 mJ/cm$^2$. The electronic temperature is very well described by the M3TM (red curves in Fig. \ref{F4}-a/b), and yields an e-ph coupling constant of G$_{ep}$ = (8.5 – 13)$\times$10$^{17}$ J K$^{-1}$ s$^{-1}$ m$^{-3}$. The electronic temperature rapidly increases within the pump pulse duration followed by e-ph relaxation within $\sim$0.5 ps, which equilibrates the electronic and lattice temperatures and leads to a transient increase of the lattice temperature by several 100 K at this fluence (dashed line in Fig. \ref{F4}-a). Subsequently, the electron and lattice temperatures relax via heat diffusion within several 100 ps (Fig. \ref{F4}-b). Remarkably, the electron and lattice temperatures remain significantly above the equilibrium surface ($T_N$ = 90 K \cite{guttler_robust_2016} exchange splitting) and bulk ($T_N$ = 107 K \cite{kliemt_single_2015,windsor_deterministic_2020} magnetic diffraction amplitude) Néel temperatures for the entire investigated time range (dashed lines in Fig. \ref{F4}-b, top).

We note here that the extracted Sommerfeld coefficient ($\gamma_0$) is at least three times larger than the reference $\gamma_0$ determined from low temperature behavior of the specific heat of GdRh$_2$Si$_2$ \cite{kliemt_single_2015}, and also depends on the fluence (Supplementary Information Section D). A similar enhancement of $\gamma_0$ necessary for describing the ultrafast electronic temperature dynamics was recently also reported for FM Ni \cite{tengdin_critical_nodate}. A possible reason could be a varying electronic density of states away from the Fermi energy \cite{li_ab_2022}. Additionally, an influence of the spin heat capacity on the electronic channel could be possible, which we considered as part of the lattice heat capacity due to the localized nature of Gd 4f spins.
Based on this description of the electronic and the lattice temperature, the dynamics of bulk and surface magnetic order are simulated. Similar to other rare-earth magnets, the M3TM features an enhanced demagnetization rate during the first $\sim$0.5 ps, corresponding to an enhanced spin-flip scattering rate due to the large transient temperature difference between electrons and lattice \cite{atxitia_ultrafast_2011}. After temperature equilibration, the demagnetization timescale slows down to $\sim$10 ps (Fig. \ref{F4}-b). According to Equation \eqref{m} of the M3TM, a recovery of magnetization is expected once the transient electron and lattice temperatures drop below the magnetic transition temperature. Remarkably, while the experimental magnetic order starts recovering at $\sim$100 ps in Fig. \ref{F4}-b, the electron and lattice temperatures stay well above the equilibrium bulk and surface Néel temperatures (dashed lines in Fig. \ref{F4}-b, top). This indicates a transiently enhanced magnetic ordering temperature, which was accounted for in the model by introducing an effective transient $T_{N}^{*}$ = 130 – 340 K (surface state exchange splitting) and 265 – 600 K (trRXD amplitude). 

While the overall behavior is well described by this model for both observables, in particular for the $\sim$10 ps timescale, a quantitative description requires a $\sim$33 \% larger value for the material-specific R factor for the surface magnetic order (see Table \ref{T1}). A possible explanation might be an underestimation of the probe penetration depth of trRXD. 

Both, the itinerant surface electrons and the localized bulk 4f moments exhibit a very similar magnetic order dynamics, described by similar microscopic physical parameters. Such a similarity of the magnetic order dynamics of the localized 4f moments and the itinerant conduction electrons has been previously considered for determining the strength of the exchange coupling between the 4f moments and conduction electrons in lanthanide-based magnets \cite{frietsch_role_nodate,frietsch_disparate_2015,rettig_itinerant_2016}. Based on such considerations, our results indicate a strong exchange coupling between the itinerant conduction electrons and the localized Gd 4f moments in AF GdRh$_2$Si$_2$. 

% One can quantify the correlation between the itinerant conduction electrons and the localized Gd 4f moments from the dynamics. From exponential fitting of the demagnetization curves of the exchange splitting and the (001) diffraction intensity, we extracted $d\Delta_{ex}/dt$ and $d\mu_{4f}/dt$, respectively, and calculated $d\Delta_{ex}/d\mu_{4f}$ = 0.0423 $\pm$ 0.0067 eV$\mu_{B}^{-1}$ at 1 mJ/cm$^2$. Compared to the equilibrium case 0.16 eV / 7.55 $\mu_{B}$ = 0.0212 eV$\mu_{B}^{-1}$, the surface conduction electrons signal reacts much stronger than the sub-surface Gd 4f signal. Why does the photoexcitation particularly enhance the correlation between the two physical parameters? What kind of implication can we give from this?

The resemblance of the magnetic order dynamics of the itinerant surface electrons and the localized bulk 4f moments allows for another interesting interpretation. Unlike the similar collinear antiferromagnet EuRh$_2$Si$_2$, where the surface state actively enhances the sub-surface in-plane FM ordering, leading to exchange splitting at significantly higher temperatures (41 K) than the bulk $T_N$ of $\sim$ 24.5 K \cite{generalov_spin_2017,guttler_divalent_2019}, GdRh$_2$Si$_2$ exhibits surface ordering at slightly lower temperatures (90 K) than the bulk ($T_N$ $\sim$ 107 K). Considering these facts, the resemblance of the two magnetic order dynamics implies that the surface state exchange splitting in GdRh$_2$Si$_2$ acts as a spectator of the sub-surface FM ordering, supporting our assignment that its dynamics can be regarded as a fingerprint of the itinerant sub-surface magnetic order.

The M3TM considers spin-flip scattering events both from Elliott-Yafet type with phonons, leading to angular momentum transfer to the lattice, as well as direct spin transfer between opposing AF sublattices \cite{jakobs_exchange-enhancement_2022}. The latter channel was recently demonstrated to contribute as an efficient demagnetization channel in GdRh$_2$Si$_2$ using trRXD and ab-initio calculations in particular for the slow demagnetization channel \cite{windsor_exchange_2022}. Therefore, we conclude that the reduction of both, the surface state exchange splitting and the 4f AF ordering, results from a combination of spin-flip scattering induced by direct spin transfer and phonon-mediated processes. This interpretation is also quantitatively supported considering the spin-flip scattering probability a$_{sf}$, which can be calculated from the material-specific R factor. For AF GdRh$_2$Si$_2$, a$_{sf}$ is 25 to 43 \% depending on the pump fluence. With the equilibrium $T_N$ = 107 K, this is two to three times larger than FM Gd (15 \%), where only a phonon-mediated process occurs \cite{koopmans_explaining_2010}. 

While the model describes the experimental results qualitatively well, the model significantly underestimates the amplitude of the fast $\sim$0.5 ps demagnetization channel (Fig. \ref{F4}-a). This is particularly evident in the exchange splitting dynamics, which exhibits a $>$10\% drop within the first 100 fs significantly exceeding the $\sim$5\% reduction in the M3TM simulations within this timescale. This could indicate an important influence of a non-thermal electron system during the first $\sim$100 fs (see Fig. \ref{F3}-a), which is neglected in the model. Such a non-thermal electron distribution could lead to more efficient spin-flip scattering due to the occupation of highly excited electronic states.

\subsection{Fluence dependence of the ultrafast magnetic order dynamics}
\begin{figure}[ht]
\centering
\includegraphics{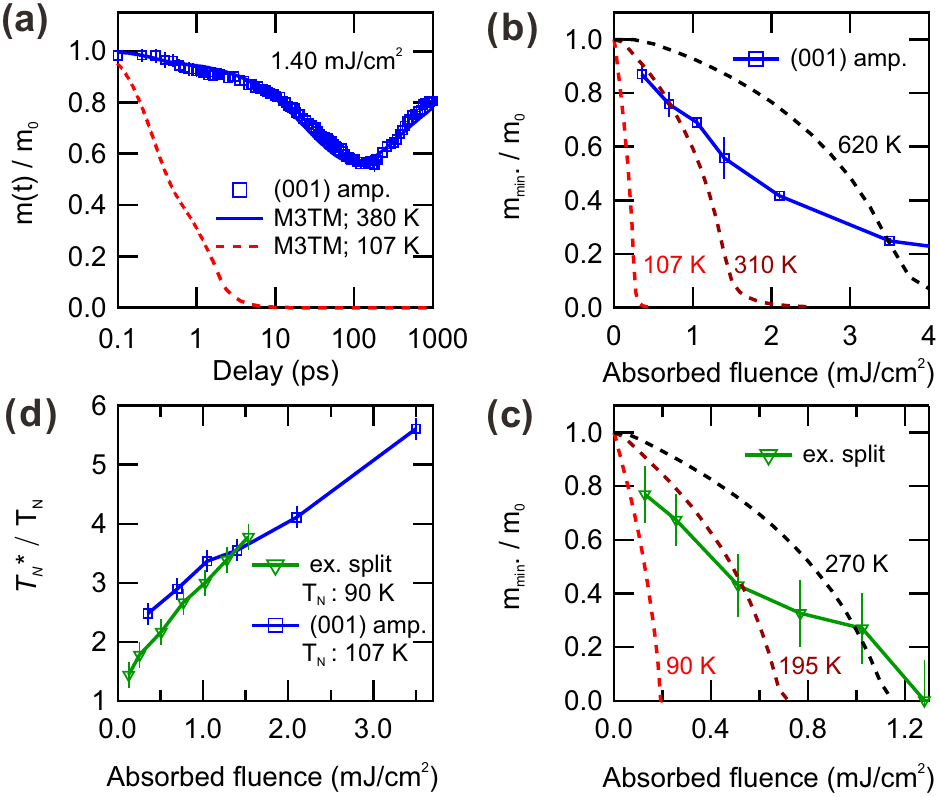}
\caption{(a) Time-dependent magnetic diffraction amplitude dynamics (blue squares) shown together with the magnetization dynamics calculated by the M3TM for two different Néel temperatures ($T_N$=107 K, red dashed line, $T_N^*$=380K, blue solid line). Note the time axis is log-scaled. (b, c) Fluence dependence of the minimal magnetic order parameter of the magnetic diffraction amplitude (b) and normalized exchange splitting (c). Dashed lines show the predictions of the M3TM for various Néel temperatures (red lines: equilibrium $T_N$) (d) Fluence dependence of the transient enhancement of $T_N^*$/$T_N$ of the diffraction amplitude (blue) and the exchange splitting (green). Note the figures are placed in clockwise order.}\label{F5}
\end{figure}
As discussed in the previous section, for a quantitative description of the ultrafast magnetic order dynamics, we have to consider a transiently enhanced transition temperature $T_N^*$. Here, we discuss the fluence dependence of this behavior. To emphasize the inability of the M3TM to account for the magnetization dynamics using the equilibrium $T_N$, Fig. \ref{F5}-a shows the trRXD amplitude at a pump fluence of 1.40 mJ/cm$^2$. Employing the equilibrium $T_N$=107 K yields a simulated magnetization dynamics exhibiting a complete demagnetization within $\sim$3 ps, which does not recover even after 1 ns. However, considering a transient $T_N^*$=380 K correctly reproduces the experimental magnetic order dynamics with $\sim$50\% demagnetization and recovery after $\sim$100 ps. Importantly, both model descriptions employ the same electronic and lattice dynamics, consistent with the experimentally measured electronic temperature (see Fig. \ref{F3}-b). Fig. \ref{F5}-b and c show the fluence dependence of the minimal normalized magnetic order parameter of the magnetic diffraction amplitude and the exchange splitting, respectively, compared to the M3TM simulations for various $T_N$. Similarly, we find that the M3TM using the equilibrium Néel temperatures cannot reproduce the experimental data. Surprisingly, even employing an enhanced critical temperature $T_N^*$ only yields a correct description of the demagnetization dynamics within a narrow fluence range, and we find that the transient $T_N^*$ scales with the pump fluence, shown in Fig. \ref{F5}-d. In other words, no single $T_N^*$ reproduces the entire experimental fluence dependence in the M3TM. Importantly, the transient Néel temperature enhancement is consistent between the trRXD data and the exchange splitting, and the electronic and lattice temperatures after e-ph equilibration following the expected fluence dependence (see Supplementary Information Section D).

Similar phenomena were previously reported for other lanthanides, however without a systematic analysis. Refs. \cite{andres_separating_2015} and \cite{andres_role_2021} reported that the exchange splitting of the 5d surface state of FM Gd and Tb starts recovering already for an electronic temperature still above the equilibrium Curie temperature. Furthermore, Ref. \cite{thielemann-kuhn_ultrafast_2017} although not directly providing a transient temperature dynamics, concluded that photoexcitation strong enough to heat the sample above the transition temperature does not lead to a complete suppression of magnetic order in FM and AF Dy. 

The fluence dependence of the minimal magnetization amplitude found in the M3TM exhibits a convex behavior, i.e. it resembles the equilibrium order parameter, with fluence acting as temperature (dashed curves in Fig. \ref{F5}-b/c). This can be understood from basic assumptions of the M3TM, that the magnetic system is not very far from thermal equilibrium. This means that the non-equilibrium magnetization is described by a thermal distribution in a non-equilibrium field, leading to magnetization dynamics governed by the equilibrium properties of the system. In contrast, the experimental data exhibit a more gradual linear to concave behavior. This behavior could indicate the importance of non-thermal spin dynamics or a transient modification of the exchange interaction in the excited system, which go beyond the currently available M3TMs. Therefore, even though such models can \emph{qualitatively} well describe experimental ultrafast magnetic order dynamics \cite{koopmans_explaining_2010,shim_role_2020,atxitia_controlling_2014,chen_ultrafast_2019,frietsch_role_nodate,sultan_electron-_2012,radu_transient_2011,frietsch_disparate_2015,atxitia_evidence_2010,nieves_quantum_2014} our results indicate the \emph{quantitative} comparison needs to be met with caution, in particular if not all experimental parameters are well known. Our data can serve as a test case for more microscopic descriptions such as atomistic spin models \cite{zahn_lattice_2021,zahn_intrinsic_2022} or time-dependent density functional theory \cite{krieger_laser-induced_2015,golias_ultrafast_2021} that go beyond a mean-field description and can potentially include such effects.

\section{Conclusion}\label{Conclusion}
In this study, employing time- and angle-resolved photoelectron spectroscopy and time-resolved resonant soft x-ray diffraction, we explored the femtosecond dynamics of the electronic temperature, the exchange splitting of a Si-derived surface state, and the resonant magnetic diffraction amplitude of the (001) magnetic reflection in antiferromagnetic GdRh$_2$Si$_2$. Combining experimental techniques sensitive to in-plane surface ferromagnetic order and out-of-plane bulk antiferromagnetic order allows us to gain a multi-faceted view on the ultrafast dynamics of magnetic order of a quasi-2D antiferromagnet. The similar dynamics of the two observables suggests a strong exchange coupling between itinerant conduction electrons, and localized Gd 4f moments. We found similar dynamics of the exchange splitting and the diffraction amplitude, which can be \emph{qualitatively} well described by a magnetic three-temperature model (M3TM) based on the Landau-Lifshitz-Bloch equation. Surprisingly, we found a recovery of the transient magnetic order already for electronic and lattice temperatures exceeding the equilibrium transition temperatures. This implies a transiently enhanced $T_{N}^{*}$, which allows us to quantitatively describe the magnetic order dynamics within the M3TM. Comparison with the mean-field behavior predicted by the M3TM suggests that the system transiently strongly deviates from a mean-field behavior. These deviations, which could be due to non-thermal effects in the spin system or a transient modification of the exchange interaction, are found to scale with increasing excitation fluence. Our results thus imply that models beyond a M3TM descriptions are necessary to quantitatively describe the ultrafast magnetic order dynamics.

% Experimental section

\section{Experimental Section}\label{Experimental Section}
%\threesubsection{Antiferromagnetic GdRh$_2$Si$_2$}
%GdRh$_2$Si$_2$ is a quasi-2D intermetallic material with a layered tetragonal crystal structure of ThCr$_2$Si$_2$ type (a = b = 4.03 Å, c = 9.88 Å), where Gd atomic layers are separated by strongly bonded Si-Rh-Si blocks along the [001] direction as shown in Fig. \ref{F1}-a \cite{kliemt_single_2015}. Below $T_N$ of 107 K, GdRh$_2$Si$_2$ undergoes a transition into an AF state. Its magnetic structure consists of ferromagnetic Gd layers in the ab-planes aligned antiferromagnetically to each other along the c-axis, i.e. along the [001] direction \cite{kliemt_mathrmgdrh_2mathrmsi_2_2017,sichelschmidt_weak_2018}. The following experiments were conducted at 20 K.\\
\threesubsection{Time- and angle-resolved photoemission spectroscopy}
Time-resolved ARPES measurements were performed using a high-repetition rate extreme ultraviolet (XUV) trARPES setup with a hemispherical analyzer \cite{puppin_500_2015}. We used 35 fs-long laser pulses centered at 1.55 eV to excite the sample, and probed the transient electronic structure by 20 fs long, 21.7 eV XUV laser pulses at a repetition rate of 500 kHz with an energy resolution of 150 meV (Fig. \ref{F1}-a). The trARPES measurements were performed on a Si-terminated surface \cite{guttler_robust_2016} of in-situ cleaved samples along the $\overline{\textrm{M}}-\overline{\textrm{X}}$ direction of the surface Brillouin zone (Fig. \ref{F1}-b), at a base pressure of $<$ 5$\times$10$^{-11}$ mbar. \\
\threesubsection{Time-resolved resonant soft x-ray diffraction}
Time-resolved resonant soft x-ray diffraction experiments were performed at the FemtoSpeX beamline of BESSY II in Berlin, Germany, which uses the femtosecond slicing technique to provide ultrashort soft x-ray pulses \cite{holldack_femtospex_2014}. 50 fs-long laser pulses centered at 1.55 eV, and at repetition rate of 3 kHz were used to excite the sample. Time-delayed, soft x-ray pulses tuned to the Gd M$_5$ absorption edge (h$\nu$ = 1.18 keV; 3d $\rightarrow$ 4f) with a pulse duration of 100 fs were used to measure the transient diffraction intensity at a repetition rate of 6 kHz with an avalanche photodiode (APD) covered by an aluminum foil to prevent detection of pump photons (Fig. \ref{F1}-a). We measured the resonantly enhanced (001) magnetic diffraction intensity (Fig. \ref{F1}-c), which vanished in the high-temperature paramagnetic state. The penetration depth of the soft x-ray light at the Gd M$_5$ threshold was estimated to $\sim$4.2 nm \cite{windsor_exchange_2022}. Further details of the RXD experiment can be found in Refs. \cite{thielemann-kuhn_ultrafast_2017,fink_resonant_2013}.

\medskip
\textbf{Supporting Information} \par %Please delete the Suppporting Information statement if it is not applicable. Please supply Supporting Information in another file. Supporting information should not be provided in .tex format
Supporting Information will follow after the References.

% Acknowledgements
\medskip
\textbf{Acknowledgements} \par %delete if not applicable))
We thank the Helmholtz-Zentrum Berlin für Materialien und Energie for the allocation of synchrotron radiation beamtime. The experimental support of the staff at beamlines UE56/1 (HZB), X11MA (SLS), and PM3 (HZB) is gratefully acknowledged. This work received funding from the Deutsche Forschungsgemeinscahft (DFG, German Research Foundation) within the Emmy Noether program (Grant No. RE 3977/1), within the Transregio TRR 227 - 328545488 Ultrafast Spin Dynamics (Projects A03, A08, and A09), within TRR 288 - 422213477 (Project A03), SFB1143 (No. 247310070) and within Grant No. 282 KR3831/5-1. Funding was also received from the European Research Council (ERC) under the European Union’s Horizon 2020 research and innovation program (Grant Agreement Number ERC-2015-CoG-682843). We also thank support from the Spanish Ministry of Science and Innovation, project PID2020-116093RB-C44, funded by MCIN/ AEI/10.13039/501100011033.

\medskip
\textbf{Conflict of Interest}
The authors declare no competing interests.

% References
\medskip
%\textbf{References}\\
% Use the following code if you wish to generate your bibliography with BibTeX;
% replace the string "MSP-template" below with the name(s) of
% the BibTeX data base(s) you want to use.
% The resulting bibliography-output (the content of the .bbl file)
% must be pasted back into this file before submission.
% Please also include your BibTeX data base file(s) in your submission
% so that we can re-run BibTeX if necessary.
%
\bibliographystyle{MSP}
\bibliography{main.bib}

\begin{thebibliography}{10}
\providecommand{\url}[1]{\texttt{#1}}
\providecommand{\urlprefix}{URL }

\bibitem{beaurepaire_ultrafast_1996}
E.~Beaurepaire, J.-C. Merle, A.~Daunois, J.-Y. Bigot,
\newblock \emph{Physical Review Letters} \textbf{1996}, \emph{76}, 22 4250.

\bibitem{koopmans_explaining_2010}
B.~Koopmans, G.~Malinowski, F.~Dalla~Longa, D.~Steiauf, M.~Fähnle, T.~Roth,
  M.~Cinchetti, M.~Aeschlimann,
\newblock \emph{Nature Materials} \textbf{2010}, \emph{9}, 3 259.

\bibitem{shim_role_2020}
J.-H. Shim, A.~A. Syed, J.-I. Kim, H.-G. Piao, S.-H. Lee, S.-Y. Park, Y.~S.
  Choi, K.~M. Lee, H.-J. Kim, J.-R. Jeong, J.-I. Hong, D.~E. Kim, D.-H. Kim,
\newblock \emph{Scientific Reports} \textbf{2020}, \emph{10}, 1 6355.

\bibitem{atxitia_controlling_2014}
U.~Atxitia, J.~Barker, R.~W. Chantrell, O.~Chubykalo-Fesenko,
\newblock \emph{Physical Review B} \textbf{2014}, \emph{89}, 22 224421.

\bibitem{chen_ultrafast_2019}
Z.~Chen, S.~Li, S.~Zhou, T.~Lai,
\newblock \emph{New Journal of Physics} \textbf{2019}, \emph{21}, 12 123007.

\bibitem{frietsch_role_nodate}
B.~Frietsch, A.~Donges, R.~Carley, M.~Teichmann, J.~Bowlan, K.~Döbrich,
  K.~Carva, D.~Legut, P.~M. Oppeneer, U.~Nowak, M.~Weinelt,
\newblock \emph{Science Advances} \textbf{2020}, \emph{6}, 39 eabb1601.

\bibitem{sultan_electron-_2012}
M.~Sultan, U.~Atxitia, A.~Melnikov, O.~Chubykalo-Fesenko, U.~Bovensiepen,
\newblock \emph{Physical Review B} \textbf{2012}, \emph{85}, 18 184407.

\bibitem{radu_transient_2011}
I.~Radu, K.~Vahaplar, C.~Stamm, T.~Kachel, N.~Pontius, H.~A. Dürr, T.~A.
  Ostler, J.~Barker, R.~F.~L. Evans, R.~W. Chantrell, A.~Tsukamoto, A.~Itoh,
  A.~Kirilyuk, T.~Rasing, A.~V. Kimel,
\newblock \emph{Nature} \textbf{2011}, \emph{472}, 7342 205.

\bibitem{frietsch_disparate_2015}
B.~Frietsch, J.~Bowlan, R.~Carley, M.~Teichmann, S.~Wienholdt, D.~Hinzke,
  U.~Nowak, K.~Carva, P.~M. Oppeneer, M.~Weinelt,
\newblock \emph{Nature Communications} \textbf{2015}, \emph{6}, 1 8262.

\bibitem{atxitia_evidence_2010}
U.~Atxitia, O.~Chubykalo-Fesenko, J.~Walowski, A.~Mann, M.~Münzenberg,
\newblock \emph{Physical Review B} \textbf{2010}, \emph{81}, 17 174401.

\bibitem{atxitia_fundamentals_2016}
U.~Atxitia, D.~Hinzke, U.~Nowak,
\newblock \emph{Journal of Physics D: Applied Physics} \textbf{2016},
  \emph{50}, 3 033003.

\bibitem{atxitia_ultrafast_2011}
U.~Atxitia, O.~Chubykalo-Fesenko,
\newblock \emph{Physical Review B} \textbf{2011}, \emph{84}, 14 144414.

\bibitem{nieves_quantum_2014}
P.~Nieves, D.~Serantes, U.~Atxitia, O.~Chubykalo-Fesenko,
\newblock \emph{Physical Review B} \textbf{2014}, \emph{90}, 10 104428.

\bibitem{zahn_lattice_2021}
D.~Zahn, F.~Jakobs, Y.~W. Windsor, H.~Seiler, T.~Vasileiadis, T.~A. Butcher,
  Y.~Qi, D.~Engel, U.~Atxitia, J.~Vorberger, R.~Ernstorfer,
\newblock \emph{Physical Review Research} \textbf{2021}, \emph{3}, 2 023032.

\bibitem{tengdin_critical_nodate}
P.~Tengdin, W.~You, C.~Chen, X.~Shi, D.~Zusin, Y.~Zhang, C.~Gentry, A.~Blonsky,
  M.~Keller, P.~M. Oppeneer, H.~C. Kapteyn, Z.~Tao, M.~M. Murnane,
\newblock \emph{Science Advances} \textbf{2018}, \emph{4}, 3 eaap9744.

\bibitem{andres_separating_2015}
B.~Andres, M.~Christ, C.~Gahl, M.~Wietstruk, M.~Weinelt, J.~Kirschner,
\newblock \emph{Physical Review Letters} \textbf{2015}, \emph{115}, 20 207404.

\bibitem{andres_role_2021}
B.~Andres, S.~E. Lee, M.~Weinelt,
\newblock \emph{Applied Physics Letters} \textbf{2021}, \emph{119}, 18 182404.

\bibitem{gort_ultrafast_2020}
R.~Gort, K.~Bühlmann, G.~Saerens, S.~Däster, A.~Vaterlaus, Y.~Acremann,
\newblock \emph{Applied Physics Letters} \textbf{2020}, \emph{116}, 11 112404.

\bibitem{perfetti_ultrafast_2007}
L.~Perfetti, P.~A. Loukakos, M.~Lisowski, U.~Bovensiepen, H.~Eisaki, M.~Wolf,
\newblock \emph{Physical Review Letters} \textbf{2007}, \emph{99}, 19 197001.

\bibitem{vasileiadis_ultrafast_2018}
T.~Vasileiadis, L.~Waldecker, D.~Foster, A.~Da~Silva, D.~Zahn, R.~Bertoni,
  R.~E. Palmer, R.~Ernstorfer,
\newblock \emph{ACS Nano} \textbf{2018}, \emph{12}, 8 7710.

\bibitem{waldecker_electron-phonon_2016}
L.~Waldecker, R.~Bertoni, R.~Ernstorfer, J.~Vorberger,
\newblock \emph{Physical Review X} \textbf{2016}, \emph{6}, 2 021003.

\bibitem{you_revealing_2018}
W.~You, P.~Tengdin, C.~Chen, X.~Shi, D.~Zusin, Y.~Zhang, C.~Gentry, A.~Blonsky,
  M.~Keller, P.~M. Oppeneer, H.~Kapteyn, Z.~Tao, M.~Murnane,
\newblock \emph{Phys. Rev. Lett.} \textbf{2018}, \emph{121} 077204.

\bibitem{jungwirth_antiferromagnetic_2016}
T.~Jungwirth, X.~Marti, P.~Wadley, J.~Wunderlich,
\newblock \emph{Nature Nanotechnology} \textbf{2016}, \emph{11}, 3 231.

\bibitem{thielemann-kuhn_ultrafast_2017}
N.~Thielemann-Kühn, D.~Schick, N.~Pontius, C.~Trabant, R.~Mitzner,
  K.~Holldack, H.~Zabel, A.~Föhlisch, C.~Schüßler-Langeheine,
\newblock \emph{Physical Review Letters} \textbf{2017}, \emph{119}, 19 197202.

\bibitem{bai_functional_2020}
H.~Bai, X.~Zhou, Y.~Zhou, X.~Chen, Y.~You, F.~Pan, C.~Song,
\newblock \emph{Journal of Applied Physics} \textbf{2020}, \emph{128}, 21
  210901.

\bibitem{windsor_exchange_2022}
Y.~W. Windsor, S.-E. Lee, D.~Zahn, V.~Borisov, D.~Thonig, K.~Kliemt, A.~Ernst,
  C.~Schüßler-Langeheine, N.~Pontius, U.~Staub, C.~Krellner, D.~V. Vyalikh,
  O.~Eriksson, L.~Rettig,
\newblock \emph{Nature Materials} \textbf{2022}, 1--4.

\bibitem{jakobs_universal_2022}
F.~Jakobs, U.~Atxitia,
\newblock Universal criteria for single femtosecond pulse ultrafast
  magnetization switching in ferrimagnets, \textbf{2022},
\newblock \urlprefix\url{http://arxiv.org/abs/2201.09067},
\newblock Number: arXiv:2201.09067 arXiv:2201.09067 [cond-mat] version: 1.

\bibitem{jakobs_exchange-enhancement_2022}
F.~Jakobs, U.~Atxitia,
\newblock Exchange-enhancement of the ultrafast magnetic order dynamics in
  antiferromagnets, \textbf{2022},
\newblock \urlprefix\url{http://arxiv.org/abs/2206.05783},
\newblock Number: arXiv:2206.05783 arXiv:2206.05783 [cond-mat].

\bibitem{kliemt_single_2015}
K.~Kliemt, C.~Krellner,
\newblock \emph{Journal of Crystal Growth} \textbf{2015}, \emph{419} 37.

\bibitem{windsor_deterministic_2020}
Y.~W. Windsor, A.~Ernst, K.~Kummer, K.~Kliemt, C.~Schüßler-Langeheine,
  N.~Pontius, U.~Staub, E.~V. Chulkov, C.~Krellner, D.~V. Vyalikh, L.~Rettig,
\newblock \emph{Communications Physics} \textbf{2020}, \emph{3}, 1 1.

\bibitem{guttler_robust_2016}
M.~Güttler, A.~Generalov, M.~M. Otrokov, K.~Kummer, K.~Kliemt, A.~Fedorov,
  A.~Chikina, S.~Danzenbächer, S.~Schulz, E.~V. Chulkov, Y.~M. Koroteev,
  N.~Caroca-Canales, M.~Shi, M.~Radovic, C.~Geibel, C.~Laubschat, P.~Dudin,
  T.~K. Kim, M.~Hoesch, C.~Krellner, D.~V. Vyalikh,
\newblock \emph{Scientific Reports} \textbf{2016}, \emph{6}, 1 24254.

\bibitem{oloff_pump_2016}
L.-P. Oloff, K.~Hanff, A.~Stange, G.~Rohde, F.~Diekmann, M.~Bauer,
  K.~Rossnagel,
\newblock \emph{Journal of Applied Physics} \textbf{2016}, \emph{119}, 22
  225106.

\bibitem{lisowski_ultra-fast_2004}
M.~Lisowski, P.~Loukakos, U.~Bovensiepen, J.~Stähler, C.~Gahl, M.~Wolf,
\newblock \emph{Applied Physics A} \textbf{2004}, \emph{78}, 2 165.

\bibitem{li_ab_2022}
Y.~Li, P.~Ji,
\newblock \emph{Computational Materials Science} \textbf{2022}, \emph{202}
  110959.

\bibitem{rettig_itinerant_2016}
L.~Rettig, C.~Dornes, N.~Thielemann-Kühn, N.~Pontius, H.~Zabel, D.~Schlagel,
  T.~Lograsso, M.~Chollet, A.~Robert, M.~Sikorski, S.~Song, J.~Glownia,
  C.~Schüßler-Langeheine, S.~Johnson, U.~Staub,
\newblock \emph{Physical Review Letters} \textbf{2016}, \emph{116}, 25 257202.

\bibitem{generalov_spin_2017}
A.~Generalov, M.~M. Otrokov, A.~Chikina, K.~Kliemt, K.~Kummer, M.~Höppner,
  M.~Güttler, S.~Seiro, A.~Fedorov, S.~Schulz, S.~Danzenbächer, E.~V.
  Chulkov, C.~Geibel, C.~Laubschat, P.~Dudin, M.~Hoesch, T.~Kim, M.~Radovic,
  M.~Shi, N.~C. Plumb, C.~Krellner, D.~V. Vyalikh,
\newblock \emph{Nano Letters} \textbf{2017}, \emph{17}, 2 811.

\bibitem{guttler_divalent_2019}
M.~Güttler, A.~Generalov, S.~I. Fujimori, K.~Kummer, A.~Chikina, S.~Seiro,
  S.~Danzenbächer, Y.~M. Koroteev, E.~V. Chulkov, M.~Radovic, M.~Shi, N.~C.
  Plumb, C.~Laubschat, J.~W. Allen, C.~Krellner, C.~Geibel, D.~V. Vyalikh,
\newblock \emph{Nature Communications} \textbf{2019}, \emph{10}, 1 796.

\bibitem{zahn_intrinsic_2022}
D.~Zahn, F.~Jakobs, H.~Seiler, T.~A. Butcher, D.~Engel, J.~Vorberger,
  U.~Atxitia, Y.~W. Windsor, R.~Ernstorfer,
\newblock \emph{Physical Review Research} \textbf{2022}, \emph{4}, 1 013104.

\bibitem{krieger_laser-induced_2015}
K.~Krieger, J.~K. Dewhurst, P.~Elliott, S.~Sharma, E.~K.~U. Gross,
\newblock \emph{Journal of Chemical Theory and Computation} \textbf{2015},
  \emph{11}, 10 4870.

\bibitem{golias_ultrafast_2021}
E.~Golias, I.~Kumberg, I.~Gelen, S.~Thakur, J.~Gördes, R.~Hosseinifar,
  Q.~Guillet, J.~Dewhurst, S.~Sharma, C.~Schüßler-Langeheine, N.~Pontius,
  W.~Kuch,
\newblock \emph{Physical Review Letters} \textbf{2021}, \emph{126}, 10 107202.

\bibitem{puppin_500_2015}
M.~Puppin, Y.~Deng, O.~Prochnow, J.~Ahrens, T.~Binhammer, U.~Morgner, M.~Krenz,
  M.~Wolf, R.~Ernstorfer,
\newblock \emph{Optics Express} \textbf{2015}, \emph{23}, 2 1491.

\bibitem{holldack_femtospex_2014}
K.~Holldack, J.~Bahrdt, A.~Balzer, U.~Bovensiepen, M.~Brzhezinskaya, A.~Erko,
  A.~Eschenlohr, R.~Follath, A.~Firsov, W.~Frentrup, L.~Le~Guyader, T.~Kachel,
  P.~Kuske, R.~Mitzner, R.~Müller, N.~Pontius, T.~Quast, I.~Radu, J.-S.
  Schmidt, C.~Schüßler-Langeheine, M.~Sperling, C.~Stamm, C.~Trabant,
  A.~Föhlisch,
\newblock \emph{Journal of Synchrotron Radiation} \textbf{2014}, \emph{21}, 5
  1090.

\bibitem{fink_resonant_2013}
J.~Fink, E.~Schierle, E.~Weschke, J.~Geck,
\newblock \emph{Reports on Progress in Physics} \textbf{2013}, \emph{76}, 5
  056502.

\end{thebibliography}

% Figures/tables and captions

% Please provide Biographies and photos for Essays, Feature Articles, Progress Reports, Reviews, and Perspectives for those authors who should be highlighted  
% These should be at most 100 words long
% For other article types this section can be removed
% Photographs should be 40mm broad and 50 mm high

%\begin{figure}
%  \includegraphics{bio-placeholder.jpg}
%  \caption*{Biography}
%\end{figure}

%\begin{figure}
%  \includegraphics{bio-placeholder.jpg}
%  \caption*{Biography}
%\end{figure}

%\begin{figure}
%  \includegraphics{bio-placeholder.jpg}
%  \caption*{Biography}
%\end{figure}

%\begin{figure}
%  \includegraphics{bio-placeholder.jpg}
%  \caption*{Biography}
%\end{figure}

% Table of contents entry should be 50 - 60 words long
% Image should be 55 mm broad and 50 mm high or 110 mm broad and 20 mm high

\appendix
\counterwithin{figure}{section}
\section{Extracting the exchange splitting from the Si-derived surface state}\label{AppendixA}
\begin{figure}
\centering
\includegraphics{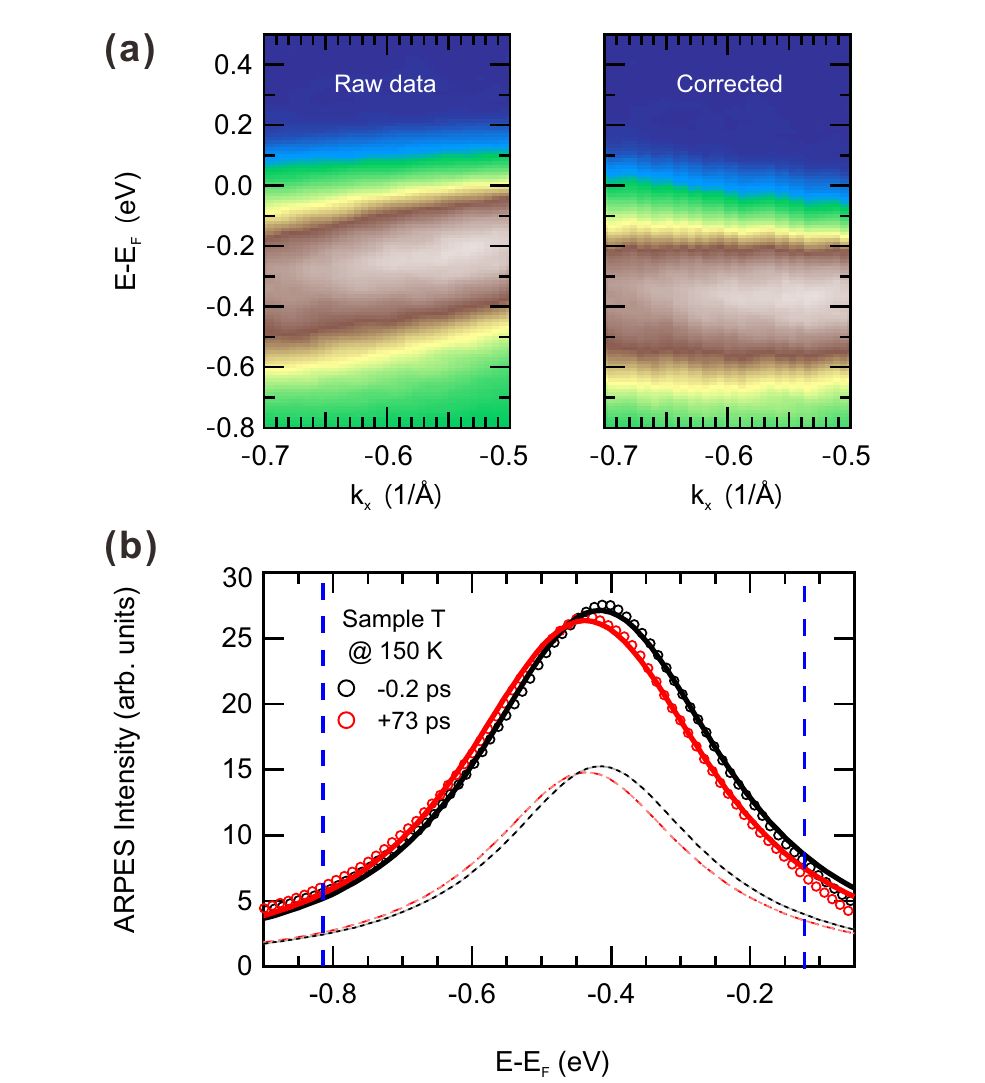}
\caption{(a) The ARPES intensity before (left) and after (right) the energy correction of the dispersion of the surface state. (b) The modeling applied for extracting the exchange splitting from EDCs measured at 150 K above T$_N$. The data well described with a vanishing exchange splitting before (black) and after (red) pump excitation. The blue dotted lines mark the region of interest used for extracting the exchange splitting.}\label{S1}
\end{figure}
The exchange-split surface state exhibits an approximately constant exchange splitting along the $\overline{\textrm{M}}-\overline{\textrm{X}}$ direction in the Brillouin zone \cite{guttler_robust_2016}. In order to utilize the statistics within the entire momentum range, we adopted a method to compensate for its energy dispersion. The energy distribution curves at each momentum have been shifted by the energy position of the surface state peak center (Fig. \ref{S1}-a, left), yielding a momentum-independent peak position (Fig. \ref{S1}-a, right). These corrected surface state data were integrated along the momentum axis between 0.5 – 0.7 1/Å to yield the EDCs shown in Fig. 2-d and \ref{S1}-b.

In order to extract the exchange splitting of the Si-derived surface state, we described the integrated EDCs with two identical Lorentzian profiles with variable exchange splitting $\Delta_{ex}$. While we find a pump-probe dependent $\Delta_{ex}$ below T$_N$ (Fig. 2-d), data taken at T=150K above T$_N$ show a vanishing exchange splitting for all pump-probe delays within error bars (Fig. \ref{S1}-b), confirming the viability of our analysis. 
\section{Space charge correction}\label{AppendixB}
\begin{figure}
\centering
\includegraphics{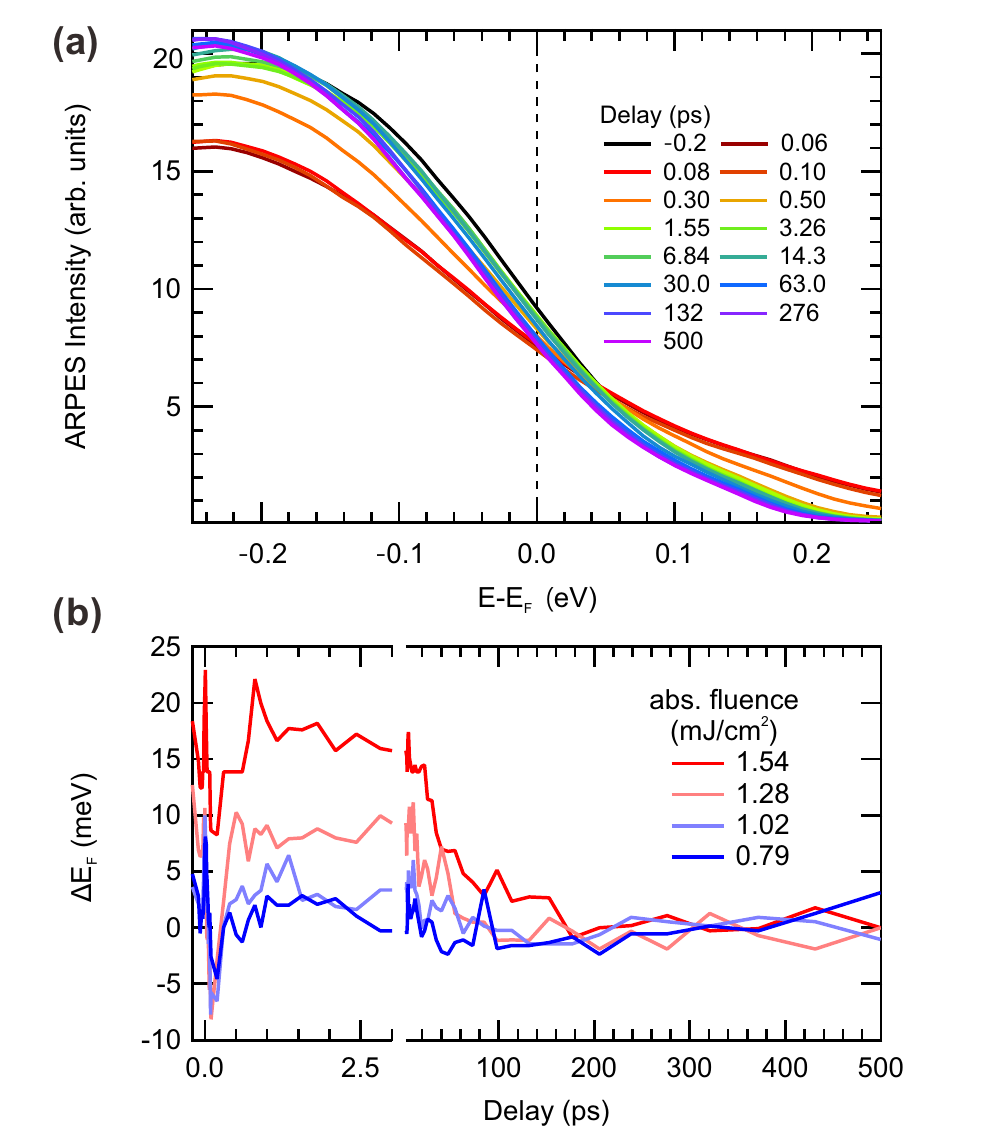}
\caption{(a) EDC at the Fermi momentum used to extract the electronic temperature for selected pump-probe delays. (b) Temporal evolution of the space-charge induced energy shift. Note that the second half of the time axis is log-scaled.}\label{S2}
\end{figure}
Upon strong pump excitation, pump-induced photoelectrons emitted from the sample lead to a broadening and energy shift of the photoelectron spectra due to space charge effects. Due to the varying distance of pump- and probe-induced electron clouds with pump-probe delays, the strength of this effect becomes time-dependent, with its strongest influence right after pump excitation, and a reduction within $\sim$100 ps \cite{oloff_pump_2016}. In order to account for this effect, we corrected this pump-induced space charge energy-shift in the EDCs by shifting E$_F$ of the Fermi-Dirac distribution of the data shown in Fig. \ref{S2}-a. The amount of time-dependent space charge shift (Fig. \ref{S2}-b) is strongly fluence dependent due to the strongly non-linear photoemission from the pump pulse, and the timescale of its reduction matches literature reports \cite{oloff_pump_2016} confirming the assignment to pump-induced space charge. 

\section{Accuracy of the electronic temperature determination}\label{AppendixC}
\begin{figure*}
\includegraphics[width=\linewidth]{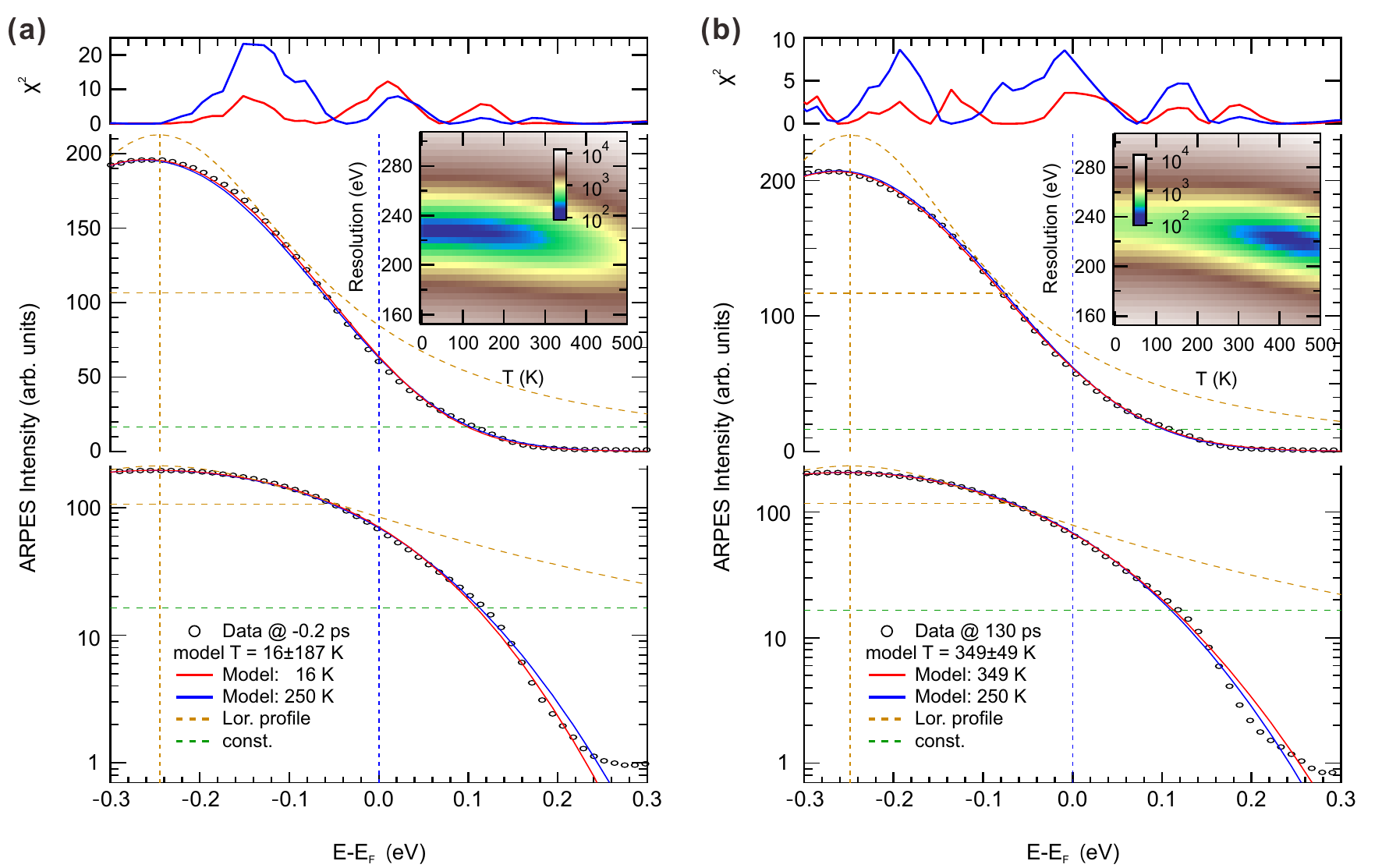}
\caption{Energy distribution curves near the Fermi momentum at (a) -0.2 ps and (b) +130 ps shown on a linear- (middle) and logarithmic intensity scale (bottom). The phenomenological density of states function used to fit the data consisting of a Lorentzian peak (yellow) and a constant offset (green) are shown as dashed lines. Fit functions at the optimized electronic temperature and at 250 K are shown as red and blue solid lines, respectively. Energy-dependent $\chi^2$ highlighting the deviations between trARPES intensity and model fits is shown in the top panel. Insets: Color-coded $\chi^2$ maps as a function of energy broadening and electronic temperature.}\label{S3}
\end{figure*}

Our assignment of a remagnetization during a transient electron and lattice temperature exceeding the equilibrium Néel temperature requires a reliable determination of the electronic temperatures, in particular at late pump-probe delays. As our limited energy resolution poses a challenge to accurately extract low electronic temperatures (compare EDCs in Fig. 3-a), we investigated the accuracy of our analysis carefully. Fig. \ref{S3} shows the EDCs and fit functions based on the same phenomenological density of states at -0.2 ps and at +130 ps along with the energy-dependent squared residual curves indicating the difference between the model and the data. The employed phenomenological density of states function consists of a Lorentzian profile (yellow dotted lines) and a constant offset (green dotted lines), multiplied by the Fermi-Dirac distribution function, and convolved with a Gaussian resolution function of a full-width at half-maximum of 230 meV. The additional broadening compared to the intrinsic energy resolution given by the spectrometer and the probe pulse bandwidth can be explained by contamination of the surface, leading to a broadening of the surface state and Fermi level during the cause of the experiment. The pump space-charge-induced broadening on the spectra is in the range of the energy shift, and negligible compared to the energy resolution.

To check the sensitivity of our fit analysis, we compare fit functions at the optimized temperatures and at 250 K in Fig. \ref{S3}. For the EDC before the photoexcitation, the higher electronic temperature (blue) yields a significant increase of the squared residual compared to the optimized temperature (16$\pm$187 K), yielding an upper limit for the electronic temperature (Fig. \ref{S3}-a). For the EDC at late times, where the magnetic order starts recovering, the situation is opposite, and the fit with optimized temperature (349$\pm$49 K) describes the data significantly better than the 250 K case (Fig. \ref{S3}-b), providing a lower limit for the electronic temperature.

To further assess the influence of the resolution function in the model, we calculated $\chi^2$ map as a function of the energy resolution and the electronic temperature parameters shown in the insets of Fig. \ref{S3}-a/b. The $\chi^2$ map before photoexcitation shows an extended minimum at the determined energy resolution of 230 meV extending until $\sim$200 K, confirming the error assignment based on the nonlinear fit parameters. After e-ph equilibration we find a slight correlation of the electronic temperature and resolution parameter, but can still clearly identify a minimum around the optimized temperature within $\pm$50 K, and with a consistent resolution function. Note also, that a reduced resolution at late times, as could be expected from reduced space charge influence, would lead to a further increase of extracted electronic temperatures, and that any description with T$_e$ $<$ T$_N$ yields a significantly worse $\chi^2$.

\section{Fluence-dependence of the electronic, spin and lattice heat capacities}\label{AppendixD}
\begin{figure}
\centering
\includegraphics{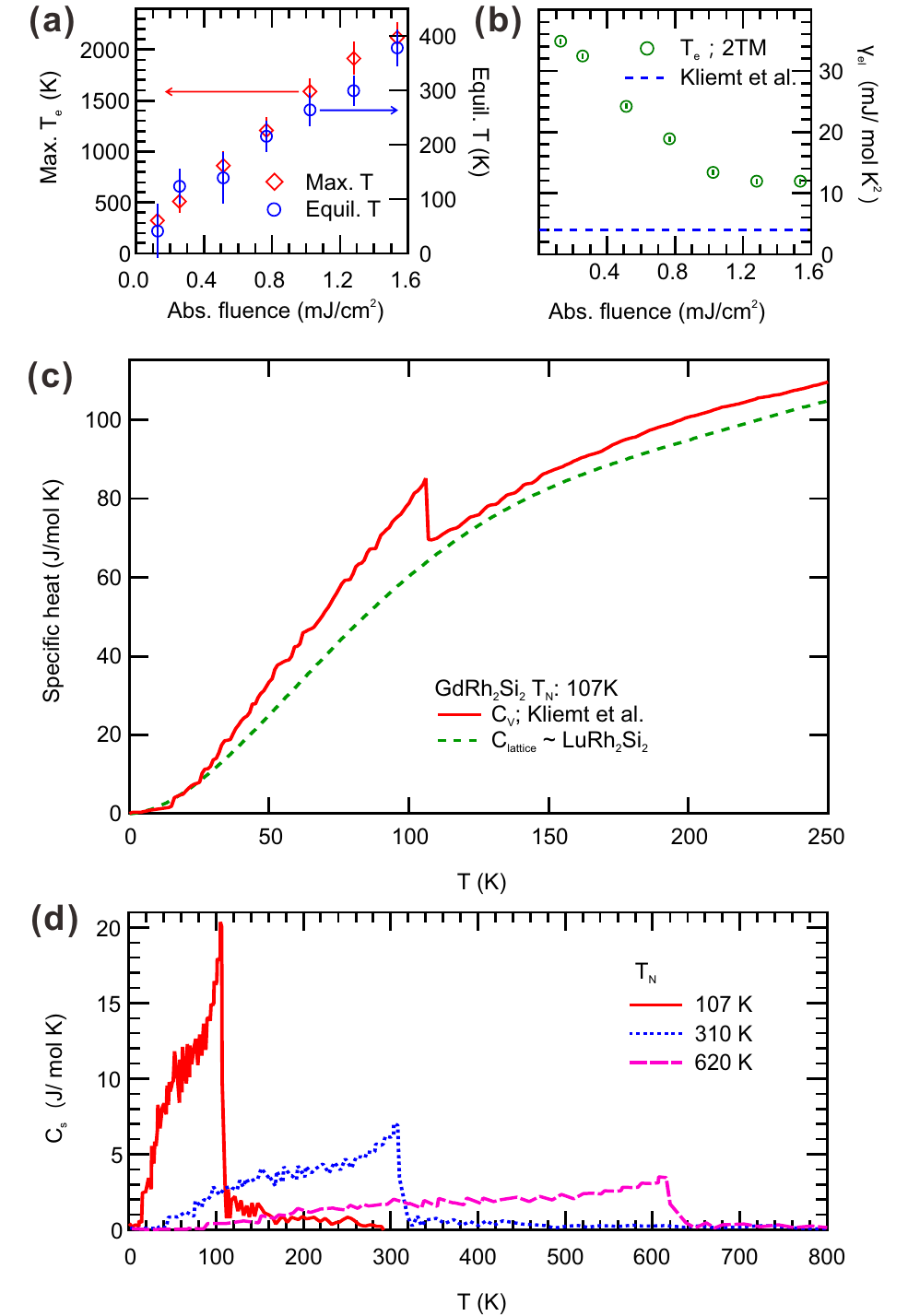}
\caption{Fluence dependence of (a) the maximum electronic temperature, the equilibrated electronic temperature and (b) the Sommerfeld coefficient, $\gamma_0(=$ $C_e/T_e)$. (c) Temperature dependence of the specific heat of GdRh$_2$Si$_2$\cite{kliemt_single_2015} and the modeled lattice heat capacity (see text). (d) Spin heat capacities of AF GdRh$_2$Si$_2$ scaled to various Néel temperatures used for the M3TM simulations described in the main text.}\label{S4}
\end{figure}
%Permission issue of reproduction of the GdRh2Si2 specific heat

The maximum electronic temperature of the M3TM is largely determined by the Sommerfeld coefficient $\gamma_0= C_e/T_e$. While the M3TM with a constant  $\gamma_0$ yields a square-root-like dependence of maximal electronic temperature with fluence, experimentally we find an approximately linear behavior (Fig. \ref{S4}-a). A consistent description of this behavior requires a fluence-dependent Sommerfeld coefficient, as shown in Fig. \ref{S4}-b. Surprisingly, all values for $\gamma_0$ that we find significantly exceed the low temperature equilibrium value reported in Ref. \cite{kliemt_single_2015}. A possible explanation could be that we account for the magnetic specific heat exclusively in the lattice system, and neglect possible contributions of itinerant magnetic specific heat to the electronic specific heat. Such an explanation would also fit to the observed fluence dependence, as the influence of the spin-induced heat capacity becomes exceedingly small compared to the electronic heat capacity as fluence and electronic temperatures increase. Please note that this description still yields a consistent modeling of the electronic and lattice temperatures, and that equilibrated electron/lattice temperatures are well-described with the modeled lattice heat capacity (Fig. \ref{S4}-a), which is a hybridization of a polynomial approximation of LuRh$_2$Si$_2$ specific heat (T $<$ 0.67 T$_D$) and the Einstein model of the lattice heat capacity of T$_D$ = 430 K (T $>$ 0.67 T$_D$) (Fig. \ref{S4}-c). 

Magnetic specific heat contributions in the M3TM were extracted from the difference of isostructural, paramagnetic LuRh$_2$Si$_2$ and AF GdRh$_2$Si$_2$ (Fig. \ref{S4}-c). For consistent M3TM simulations, the spin heat capacity was scaled to the transient Néel temperature (see Fig. \ref{S4}-d) and considered as part of the lattice heat capacity. Please note that M3TM simulations of the electron and lattice temperatures do hardly depend on T$_N^*$, and that this correction only marginally influences the exact values of fitting parameters. \\

\begin{figure}
\centering
\textbf{Table of Contents}\\
\medskip
  \includegraphics{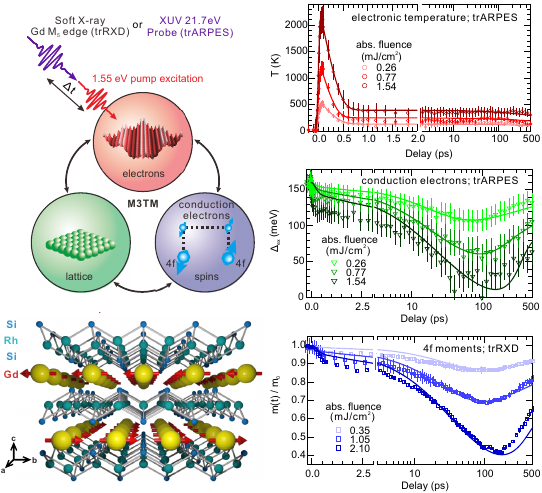}
  \medskip
  \caption*{Femtosecond dynamics of electronic temperature, sub-surface ferromagnetic ordering and bulk 4f antiferromagnetic ordering of antiferromagnetic GdRh$_2$Si$_2$ upon optical excitation were explored at various pump fluences. For consistent description of the demagnetization dynamics, the microscopic three temperature model was applied. While the model qualitatively describes the experimental results well, quantitatively, it suggests a systematic effective increase of the phase transition temperature.}
\end{figure}

\end{document}